%% file: main.tex
\documentclass[sigconf]{acmart}

\AtBeginDocument{%
  }

\setcopyright{acmlicensed}
\copyrightyear{2026}
\acmYear{2026}
\acmConference[Preprint '26]{Preprint}{2026}{}
\acmBooktitle{{Preprint}}

\usepackage{xcolor}
\usepackage{xspace}
\usepackage{multirow}
\usepackage{fontawesome5}

\newcommand{\beir}{\textsc{BEIR}\xspace}
\newcommand{\vidore}{\textsc{ViDoRe}\xspace}
\newcommand{\msrvtt}{\textsc{MSR-VTT}\xspace}
\newcommand{\multivent}{\textsc{MultiVENT 2.0}\xspace}
\newcommand{\seq}{\textsc{SeqResize}\xspace}
\newcommand{\mem}{\textsc{MemTok}\xspace}
\newcommand{\ours}{\textsc{AGC}\xspace}
\newcommand{\hcluster}{\textsc{H-Pool}\xspace}

\begin{document}

\title{Multi-Vector Index Compression in Any Modality}


\author{Hanxiang Qin}
\email{hqin14@jhu.edu}
\affiliation{%
  \institution{Johns Hopkins University}
  \city{Baltimore}
  \state{MD}
  \country{USA}}
\authornote{Equal Contribution}
  
\author{Alexander Martin}
\email{amart233@jhu.edu}
\affiliation{%
  \institution{Johns Hopkins University}
  \city{Baltimore}
  \state{MD}
  \country{USA}}
\authornotemark[1]

\author{Rohan Jha}
\email{rjha5@jhu.edu}
\affiliation{%
  \institution{Johns Hopkins University}
  \city{Baltimore}
  \state{MD}
  \country{USA}}

\author{Chunsheng Zuo}
\email{czuo3@jhu.edu}
\affiliation{%
  \institution{Johns Hopkins University}
  \city{Baltimore}
  \state{MD}
  \country{USA}}

\author{Reno Kriz}
\email{rkriz1@jhu.edu}
\affiliation{%
  \institution{Johns Hopkins University}
  \city{Baltimore}
  \state{MD}
  \country{USA}}

\author{Benjamin Van Durme}
\email{vandurme@jhu.edu}
\affiliation{%
  \institution{Johns Hopkins University}
  \city{Baltimore}
  \state{MD}
  \country{USA}}

\renewcommand{\shortauthors}{Qin et al.}

\begin{abstract}
  \input{sections/00-abstract}
\end{abstract}




\maketitle

\input{figures/teaser}

\section{Introduction}
\label{section:intro}
\input{sections/10-intro}

\section{Related Work}
\label{section:related}
\input{sections/20-related}

\section{Preliminaries}
\label{section:prelim}
\input{sections/30-prelim}

\section{Multi-Vector Compression}
\label{section:method}

\input{sections/40-compression}

\section{Experiments}
\label{section:experiments}
\input{sections/60-experiments}

\section{Conclusion}
\label{section:conclusion}

\input{sections/70-conclusion}

\begin{acks}
  This material is based upon work supported by the NSF Graduate Research Fellowship under Grant No. DGE2139757. Any opinion, findings, and conclusions or recommendations expressed in this material are those of the author(s) and do not necessarily reflect the views of the National Science Foundation.
\end{acks}

\clearpage


\bibliographystyle{ACM-Reference-Format}
\bibliography{main}




\end{document}

%% file: sections/00-abstract.tex
We study efficient multi-vector retrieval for late interaction in any modality. Late interaction has emerged as a dominant paradigm for information retrieval in text, images, visual documents, and videos, but its computation and storage costs grow linearly with document length, making it costly for image-, video-, and audio-rich corpora. To address this limitation, we explore query-agnostic methods for compressing multi-vector document representations under a constant vector budget. We introduce four approaches for index compression: sequence resizing, memory tokens, hierarchical pooling, and a novel attention-guided clustering (\ours). \ours uses an attention-guided mechanism to identify the most semantically salient regions of a document as cluster centroids and to weight token aggregation. Evaluating these methods on retrieval tasks spanning text (\beir), visual-document (\vidore), and video (\msrvtt, \multivent), we show that attention-guided clustering consistently outperforms other parameterized compression methods~(sequence resizing and memory tokens), provides greater flexibility in index size than non-parametric hierarchical clustering, and achieves competitive or improved performance compared to a full, uncompressed index.\footnote{The source code is available at: \href{https://github.com/hanxiangqin/omni-col-press}{\faGithub\ github.com/hanxiangqin/omni-col-press}}

%% file: figures/teaser.tex
\begin{figure}[t]
\vspace{-0.5em}
    \centering
    \includegraphics[width=\linewidth]{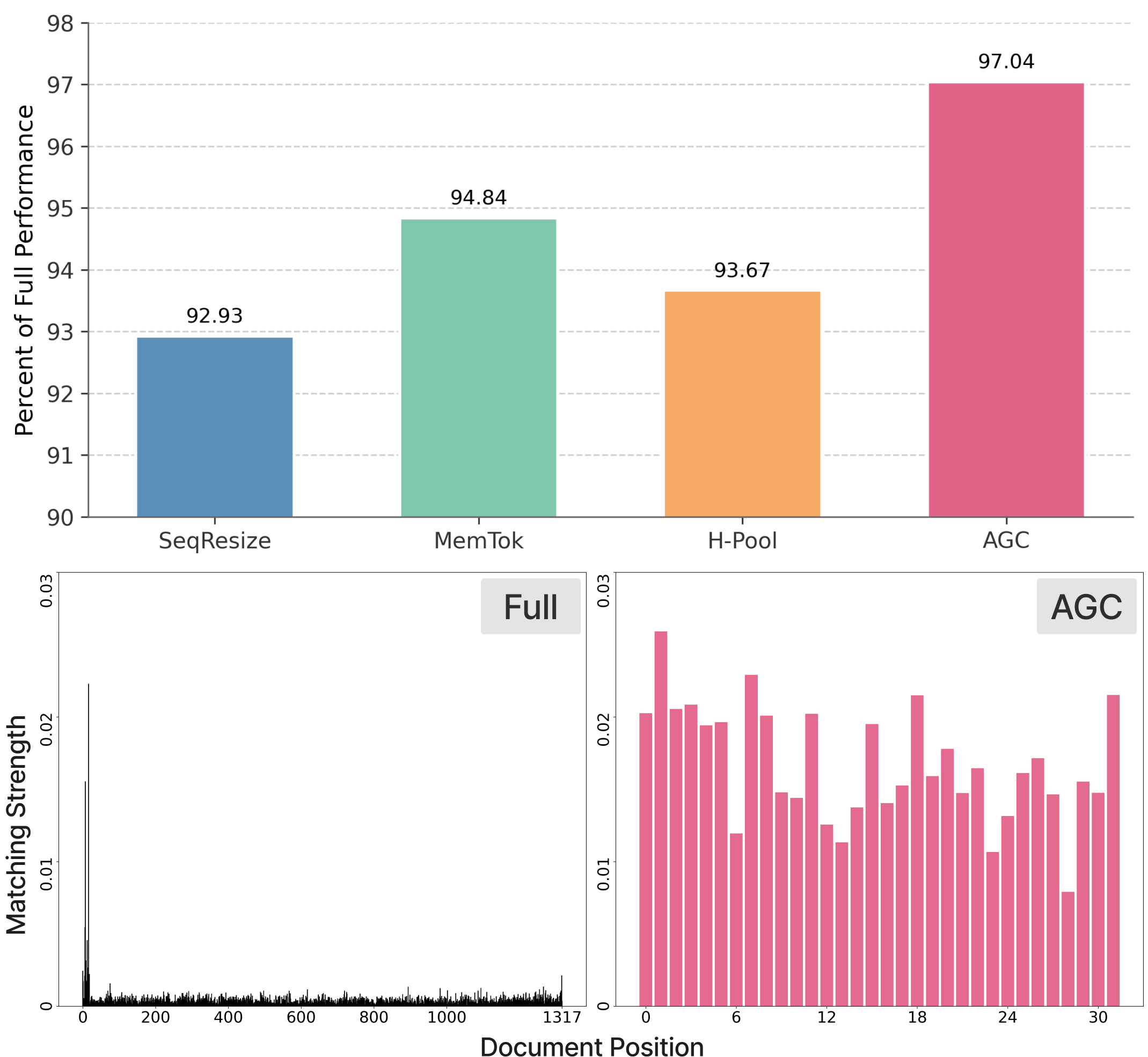}
    \caption{We explore index compression in any modality. We introduce \seq, projection-based, \mem, token-based, \hcluster, heuristic-based, and \ours (Ours), hybrid attention-similarity. \ours better utilizes index tokens while maintaining performance (nDCG@10) at high compression.}
    \label{fig:teaser}
    \vspace{-1em}
\end{figure}

%% file: sections/10-intro.tex
Online information is increasingly multimodal, including videos, articles with figures or images, podcasts, and interactive web content. It is thus essential that information retrieval systems be able to index over large multimodal collections. However, indexing multimodal content at scale requires an incredible amount of storage,\footnote{Indexing 1 video takes 10MB under multi-vector setting, but it is estimated that YouTube hosts 14 billion videos \cite{mcgrady2024deepyoutube}, estimating an index size of 140 Petabytes.} limiting the ability of search providers to build truly multimodal indices. While recent advances in multi/omnimodal retrieval~\cite{faysse2024colpaliefficientdocumentretrieval, ma2025tevatron20unifieddocument, reddy2025videocolbertcontextualizedlateinteraction, chen2026e5omniexplicitcrossmodalalignment, li2026qwen3vlembeddingqwen3vlrerankerunifiedframework} have begun to make performance gains, significant progress is still needed to achieve scalable performance in real-world settings.

In this work, we focus on multi-vector late interaction \cite{khattab2020colbertefficienteffectivepassage}, which has shown promise in multimodal domains \cite{reddy2025videocolbertcontextualizedlateinteraction, faysse2024colpaliefficientdocumentretrieval, gunther2025jinaembeddingsv4universalembeddingsmultimodal, xu2025llamanemoretrievercolembedtopperforming, moreira2026nemotroncolembedv2topperforming, wan2025clamrcontextualizedlateinteractionmultimodal}. Optimizations like ColBERTv2 \cite{santhanam2022colbertv2effectiveefficientretrieval} have improved efficiency through a two-stage retrieval pipeline that uses document cluster centroids to avoid scoring every document, enabling sub-linear scaling in collection size. However, the computation and storage cost still grow linearly with document length \cite{xiao2025metaembedscalingmultimodalretrieval, santhanam2022plaidefficientenginelate}. This linear scaling presents a prohibitive barrier for multimodal corpora, where a single multimodal document could easily reach thousands of tokens. Additionally, these thousand token representations are often underutilized in late-interaction, making the indices built from them largely unnecessary in practice (\autoref{fig:teaser}). We find that most multimodal late interaction models use only about 1\% of their index during a full evaluation pass. To address this gap, we propose learning compact, query-agnostic, multi-vector, multimodal document representations under a constant vector budget. 
By bounding the document representation to a constant size, we ensure that both index storage and query-time costs remain manageable and fully customizable to storage or compute constraints, while retaining the benefits of fine-grained late interaction.

We adapt three strong performing multi-vector compression methods from textual domain to multimodal: (1) \emph{Sequence Resizing} (\seq) \cite[e.g.,][]{macavaney2025efficientconstantspacemultivectorretrieval}, where a full multi-vector document is projected down along the sequence dimension by an MLP; (2) \emph{Memory Tokens} (\mem) \cite[e.g.,][]{louis2025piscoprettysimplecompression, xiao2025metaembedscalingmultimodalretrieval}, where learnable vectors are appended to the document context and used as the representation; and (3) \emph{Hierarchical Pooling} (\hcluster) \cite[e.g.,][]{clavie2024reducingfootprintmultivectorretrieval}, which iteratively groups similar vectors and replaces them with their mean.
However, these methods are ill-suited for multimodal compression as they struggle to handle redundant and noisy inputs, or suffer from representation collapse. To address these limitations, we introduce a novel attention-guided clustering, where learnable universal query tokens are used to guide the attention to select centroids and weight the aggregation for clustering (\ours).

We evaluate these methods across four tasks and three modalities: \beir \cite{thakur2021beirheterogenousbenchmarkzeroshot}, a document retrieval benchmark (text), \vidore \cite{mace2025vidorebenchmarkv2raising}, a visual document retrieval benchmark (vision), \msrvtt \cite{xu2016msrvttlargevideodescription}, a video-retrieval benchmark (vision), and \multivent \cite{kriz2025multivent20massivemultilingual}, a video-retrieval benchmark (audiovisual). On these benchmarks, we provide an extensive set of experiments and introduce new state-of-the-art results on \vidore and \msrvtt. We find that \ours is the strongest compression technique in any modality, offering the best performance at learned compression rates and better transferability between sizes than non-parametric compression (\hcluster). Additionally, we find that training with a compression objective can improve performance over an uncompressed multi-vector index on \vidore and \msrvtt, highlighting that compression reduces the redundancy and noise of multimodal inputs.

Our contributions are summarized as follows:
\begin{enumerate}
    \item We introduce four methods for index compression in any modality: \seq, \mem, \hcluster, and \ours.
    \item \ours presents a novel approach, in which learnable universal query tokens select centroids and weight cluster pooling. 
    \item We present a series of experiments demonstrating the strong performance and flexibility of \ours across document, visual document, and video retrieval settings.
\end{enumerate}

%% file: sections/20-related.tex
\paragraph{Multimodal Retrieval}
Many works have introduced benchmarks for evaluating representations in information retrieval. In the text-only setting, evaluation suites such as MS MARCO \cite{bajaj2018msmarcohumangenerated} and  BEIR \cite{thakur2021beirheterogenousbenchmarkzeroshot}, have become standard for measuring retrieval effectiveness across diverse domains, tasks, and query types. Video retrieval has been extensively studied using benchmarks that use natural language descriptions to retrieve videos, such as, MSR-VTT \cite{xu2016msrvttlargevideodescription}, VATEX \cite{wang2019vatexlargescalehighqualitymultilingual}, DiDeMo \cite{hendricks2017localizingmomentsvideonatural}, and ActivityNet Captions \cite{krishna2017densecaptioningeventsvideos}. More recently, MultiVENT~2.0 \cite{kriz2025multivent20massivemultilingual} has provided a large-scale, multilingual benchmark for real world video retrieval. Visual document retrieval has also recently emerged as another challenging multimodal task, requiring strong optical character recognition ability and visual understanding of layout and graphics, e.g., ViDoRe \cite{mace2025vidorebenchmarkv2raising} and MMDocIR \cite{dong2025mmdocirbenchmarkingmultimodalretrieval}. Complementary efforts have introduced modality-specific embedding benchmarks, including MTEB for text embeddings \cite{muennighoff2023mtebmassivetextembedding}, MSEB for audio embeddings \cite{heigold2026massivesoundembeddingbenchmark}, and MMEB for vision-language embeddings \cite{jiang2024vlm2vectrainingvisionlanguagemodels}. In this work, we focus on the following settings: text, visual document, video (vision only) and video (audiovisual). We believe this covers the most challenging combinations of modalities and model capabilities. 

\paragraph{Multi-Vector Index Compression}

Multi-vector embeddings offer a number of distinct axes by which to compress the index. 
Naturally, being just a collection of vectors, it is amenable to the same quantization \cite{jegou2011productquantizationnearestneighbor, fang2022jointoptimizationmultivectorrepresentation} and truncation \cite{kusupati2022matryoshka, jha2024jinacolbertv2generalpurposemultilinguallate} methods as single-vector retrieval.
It is also the norm in multi-vector text retrieval to down-project the encoder's large hidden dimension to a more manageable dimension (e.g. $768 \to 128$) \cite{khattab2020colbertefficienteffectivepassage}.
Furthermore, as is the focus of this paper, \textit{multi}-vector indices can be compressed along the sequence dimension as well in order to end up with fewer tokens. 
This is generally split between methods that prune tokens according to simple corpus-level or attentional heuristics \cite{acquavia2023staticpruningmultirepresentationdense, zong2025losslesstokenpruninglateinteraction}, pool them implicitly via special tokens that aggregate meaning or explicitly via heuristic representation merging \cite{hofstatter2022introducingneuralbagwholewords, lassance2021studytokenpruningcolbert, qin2023nuggetneuralagglomerativeembeddings, clavie2024reducingfootprintmultivectorretrieval, veneroso2025crispclusteringmultivectorrepresentations}, or simply project the sequence length into a fixed quantity of embeddings \cite{macavaney2025efficientconstantspacemultivectorretrieval}.
Finally, index methods like PLAID \cite{santhanam2022plaidefficientenginelate, santhanam2022colbertv2effectiveefficientretrieval} cluster the document token vectors and represent each as its nearest cluster centroid plus a low-bit-quantized version of its residual.

\paragraph{Attention-based Compression}

To address the computational burden of long contexts for language models, prior work explores token compression via KV cache eviction. While query-aware methods effectively prune tokens based on prompt attention~\cite{zhang2023h2oheavyhitteroracleefficient, li2024snapkvllmknowswhat, chen2024imageworth12, zhang2025sparsevlmvisualtokensparsification, huang2025prunevidvisualtokenpruning, tao2025dycokedynamiccompressiontokens, liu2024multistagevisiontokendropping}, they are incompatible with retrieval indexing, which requires document representations to be computed before the query is known. On the query-agnostic side, approaches instead leverage self-attention scores or learnable parameters to determine token importance~\cite{bolya2022tokenmergingyourvit, chari2025compactorcalibratedqueryagnostickv, yang2025visionziplongerbetternot, qin2023nuggetneuralagglomerativeembeddings, zhu2025visionselectorendtoendlearnablevisual}. However, a critical gap remains: these methods optimize for generative tasks by preserving the global "gist," whereas retrieval requires retaining discriminative details needed for distinguishing hard negatives from positive documents.

%% file: sections/30-prelim.tex
\newcommand{\D}{\mathcal{D}}

\paragraph{Retrieval}
Given a collection of documents $\D$, where each document $d \in \mathcal{D}$ contains one or more modalities (e.g., text, audio, visual), and a text query $q$, we look to provide a ranking of documents in $\D$ based on their relevance to $q$. 

\paragraph{Late Interaction}
To calculate the similarity score $\mathcal{S}$ between two multi-vector representations, we adopt ColBERT-style Late Interaction \cite{khattab2020colbertefficienteffectivepassage}. 

Given a query representation $\mathbf{Q}$ and a document representation $\mathbf{C}$, we compute the relevance score $s(q,d)$ via the MaxSim operation:
\[
  s(q,d)
  = \sum_{i=1}^{n_q} \max_{1 \le j \le m} \, \langle \mathbf{q}_i, \mathbf{c}_{j} \rangle
\]
where $\langle \cdot, \cdot \rangle$ denotes the dot product. By summing the maximum similarities for each $q_i$, we get a score for $d$'s relevance to $q$. 

\input{figures/methods}
\subsection{Problem Formulation}
We formulate the compression of multimodal documents as the process of generating optimal representations under constraints for scalable late-interaction retrieval.

\paragraph{Query and Document Representation}
We employ a query encoder $\phi$ that maps a query $q$ to a sequence of token embeddings $\mathbf{Q} \in \mathbb{R}^{n_q \times h}$, where $n_q$ is the sequence length and $h$ is the embedding dimension. We do not constrain the query length. 

For documents, we define a mapping $\pi$ that transforms a document $d$ into a sequence of $m$ vectors:
\[
  \pi: d \mapsto \mathbf{C} = \big[\,\mathbf{c}_{1}, \ldots, \mathbf{c}_{m}\,\big] \in \mathbb{R}^{m \times h}
\]
Here, $m$ is a fixed budget of vectors independent of the document's original length. The mapping $\pi$ represents the full representation generation pipeline which may 
involve direct encoding, parametric or heuristic compression, or a combination of these operations.

A critical constraint in retrieval is that this mapping must be applied during indexing, where the query $q$ remains unknown. Consequently, $\pi$ must compress the document in a query-agnostic manner while preserving information likely to be relevant for future queries. In this work, we explore both unparameterized and parameterized compression techniques, denoting parameterized formulations as $\pi_\theta$, where $\theta$ denotes learnable weights. 

\paragraph{Objective}
Our goal is to define $\pi$ such that a scoring function $\mathcal{S}(\mathbf{Q}, \mathbf{C})$ assigns higher scores to relevant query-document pairs compared to less relevant ones. For parameterized mappings, this becomes an optimization problem where we seek to maximize retrieval accuracy within the fixed storage budget $m$.

%% file: figures/methods.tex
\begin{figure*}
    \centering
    \includegraphics[width=1\linewidth]{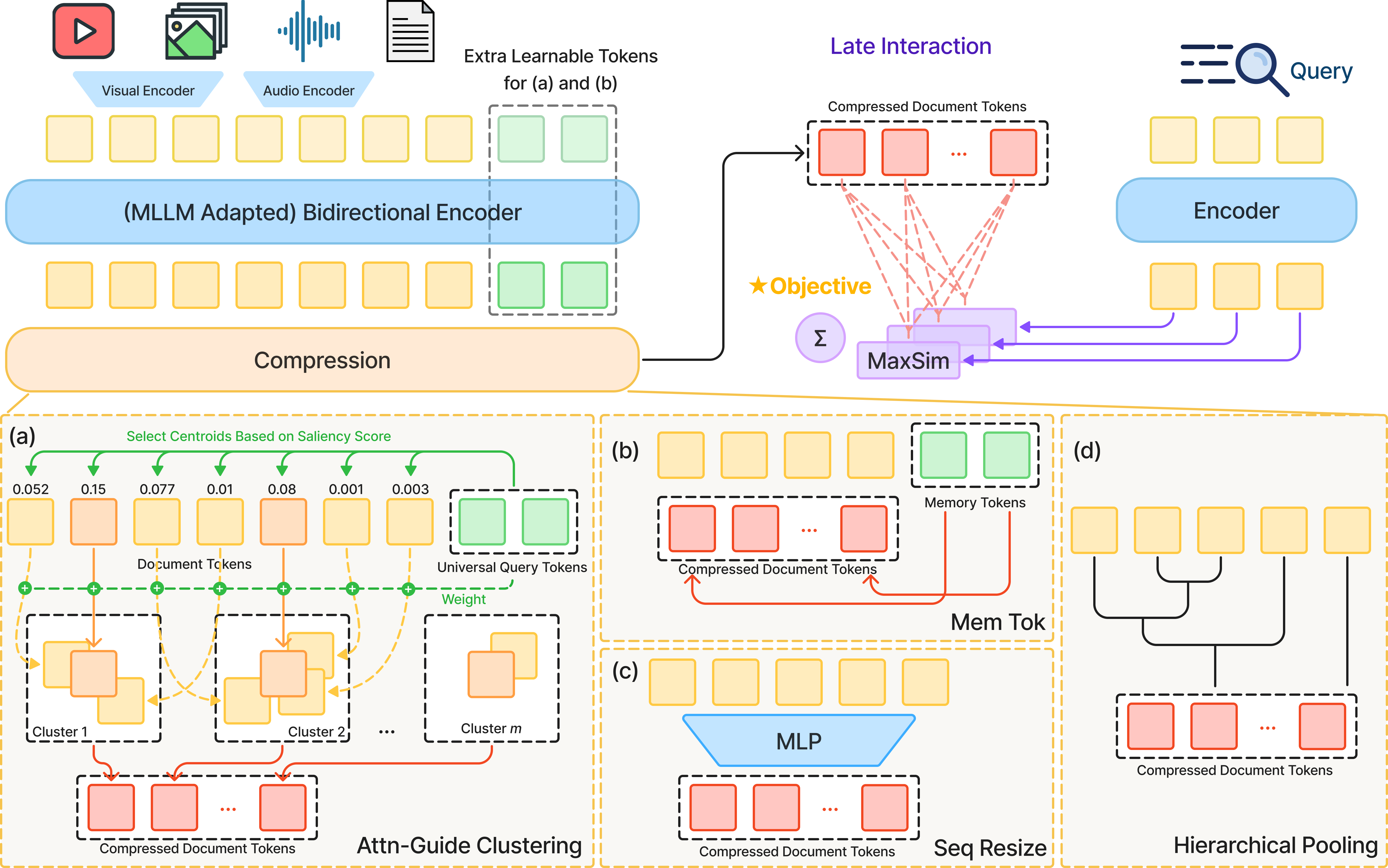}
    \caption{Overview of multi-vector index compression techniques. (a) \ours uses universal query tokens to guide attention-based centroid selection and weight the aggregation of clustering. (b) \mem appends tokens to the document context to act as the final representation. (c) \seq down projects a document representation along the sequence dimension. (d) \hcluster iteratively groups similar vectors and replaces them with their mean.}
    \label{fig:main}
\end{figure*}

%% file: sections/40-compression.tex
We introduce three methods for multi-vector compression based on prior work (pictured in \autoref{fig:main}): sequence resizing (\seq), memory tokens (\mem), and hierarchical pooling (\hcluster).

\subsection{Sequence Resizing}
\label{section:seq-resize}
\input{sections/41-resize}

\subsection{Memory Tokens}
\label{section:mem-tok}
\input{sections/42-memtok}

\subsection{Hierarchical Pooling}
\label{section:hcluster}
\input{sections/43-hcluster}

\subsection{Limitations}

\seq, \mem, and \hcluster reveal limitations when applied to multimodal data under a fixed token budget constraint. First, parametric methods like \seq exhibit a modeling failure in which many tokens remain unused during a single evaluation pass; consequently, it fails to scale effectively with the token budget (see \autoref{subsec:stability}). \mem suffers from information collapse: its architecture inherently smooths over distinct features, impeding the effective utilization of multi-vector representations. (See \autoref{subsec:util}). 
Second, neither \seq nor \mem possesses the necessary heuristics to eliminate redundant information. Audio and visual signals are often semantically empty or redundant, such as silent audio segments, static backgrounds, and unchanged temporal sequences \cite{barlow2001redundancyreductionrevisited, legall1991mpegvideocompressionstandard}.
These methods waste their limited token budgets on encoding repetitive and noisy signals rather than capturing key semantic content. 
Finally, while \hcluster actively removes redundancy, the reliance on greedy iterative merging makes them vulnerable to noisy outliers, like the aforementioned noise in multimodal data. 

\section{Attention-Guided Clustering}
\label{section:ours}
\input{sections/50-method}

%% file: sections/41-resize.tex
\seq is a parameterized compression method that projects the output of an encoder along the sequence dimension to a compressed representation with a fixed number of tokens. Prior work first used this method in document compression for text retrieval~\cite{macavaney2025efficientconstantspacemultivectorretrieval}. \seq is an intuitive approach, allowing an encoder to fully contextualize all representations in a document and then parameterize compression separately (but trained jointly) from the encoder. 

To perform \seq, the tokens of the document $\mathbf{X} \in \mathbb{R}^{n \times h}$ are passed into the bidirectional encoder $F_\text{enc}$, which is an $L$-layer transformer. Let $\mathbf{Z}^{(L)} = F_\text{enc}(\mathbf{X}; \theta) \in \mathbb{R}^{n \times h}$ be the last-layer hidden states. Since $n$ varies across documents, we first pad or truncate $\mathbf{Z}^{(L)}$ to a fixed length $n_0$:
\[
  \bar{\mathbf{Z}}^{(L)} = \text{PadTrunc}(\mathbf{Z}^{(L)}, n_0) \in \mathbb{R}^{n_0 \times h}.
\]
We then resize along the sequence dimension to produce the compressed multi-vector representation $\mathbf{C} \in \mathbb{R}^{m \times h}$ using 
a 2-layer MLP:
\begin{align*}
  \mathbf{C} &= \Big(\sigma\!\big(\bar{\mathbf{Z}}^{(L)\top}\mathbf{W}_1^\top\big)\mathbf{W}_2^\top\Big)^\top, 
  &&\mathbf{W}_1 \in \mathbb{R}^{d \times n_0},\ \mathbf{W}_2 \in \mathbb{R}^{m \times d}.
\end{align*}
Here, $h$ is the hidden dimension, $\theta$ are the parameters of the encoder, and $\sigma$ is a nonlinearity (e.g., ReLU). The transpose in the MLP form indicates that the same MLP maps each hidden channel’s length-$n_0$ sequence to a length-$m$ sequence, i.e., it operates over the sequence dimension. The parameters of 
$\theta, \mathbf{W}_1,\mathbf{W}_2$ are the learnable parameters of the compression function.

%% file: sections/42-memtok.tex
\mem is a parameterized compression method that appends learnable "memory tokens" to a document context to use as the document representation. Prior works have used this method in document compression for retrieval \cite{xiao2025metaembedscalingmultimodalretrieval} and generation \cite{louis2025piscoprettysimplecompression}. \mem approaches are a straightforward compression method, allowing for document compression to leverage a single encoder, instead of parameterizing compression in a different network (e.g., \seq). Following these works, we aim to learn memory tokens for any document representation. 

To perform compression with \mem, we append a set of $m$ memory tokens $\mathbf{M} \in \mathbb{R}^{m \times h}$ to the document tokens $\mathbf{X} \in \mathbb{R}^{n \times h}$, and feed the concatenated sequence into the encoder $F_\text{enc}$, which is an $L$-layer transformer. After bidirectional self-attention, each memory token attends over the entire document, and we discard all non-memory positions. The final states of the $m$ memory tokens form the compressed multi-vector representation of the document.

Let $\mathbf{Z}^{(L)} = [\mathbf{Z}_{\mathbf{X}}, \mathbf{Z}_{\mathbf{M}}] \in \mathbb{R}^{(n+m) \times h}$ be the hidden states of the last layer of the encoder, where $\mathbf{Z}_{\mathbf{X}} \in \mathbb{R}^{n \times h}$ and $\mathbf{Z}_{\mathbf{M}} \in \mathbb{R}^{m \times h}$ are the hidden states of the document and memory tokens, respectively. Formally, the compressed multi-vector representation of the document $\mathbf{C} \in \mathbb{R}^{m \times h}$ is given by:

\begin{align*}
  [\mathbf{Z}^{(L)}_\mathbf{X}, \mathbf{Z}^{(L)}_\mathbf{M}] &= F_\text{enc}([\mathbf{X}, \mathbf{M}]; \theta), \\
  \mathbf{C} = [\mathbf{c}_1, \mathbf{c}_2, \ldots, \mathbf{c}_m] &= \mathbf{Z}^{(L)}_\mathbf{M}.
\end{align*}

where $h$ is the hidden dimension, $\theta$ are the parameters of the encoder, which are the learnable parameters of the compression function. In our experiments, the memory tokens $\mathbf{M}$ are initialized as learnable parameters and updated during training.

%% file: sections/43-hcluster.tex
\hcluster is a non-parametric compression method that iteratively groups similar vectors and replaces them with their mean. Prior work has introduced this method in document compression in the text domain \cite{clavie2024reducingfootprintmultivectorretrieval}. Unlike the other approaches, \hcluster does not require the model to be trained for compression and allows for a simple heuristic-driven approach agnostic to modality or model. 

For each sequence, \hcluster starts from a sequence of token embeddings $\mathbf{X} \in \mathbb{R}^{n \times h}$. We compute a cosine distance matrix $\mathbf{R} \in \mathbb{R}^{n \times n}$ with entries
\[
r_{ij} = 1 - \frac{x_i^\top x_j}{\lVert x_i \rVert_2 \, \lVert x_j \rVert_2},
\]
which we use to run agglomerative hierarchical pooling with Ward linkage \cite{ward1963hierarchicalgroupingoptimizeobjective}. At any step of the algorithm, we maintain clusters as index sets $\{\mathcal{A}_1, \dots, \mathcal{A}_k\}$; each cluster $\mathcal{A}_a$ has centroid
\[
\mu_a = \frac{1}{|\mathcal{A}_a|} \sum_{i \in \mathcal{A}_a} x_i.
\]
Ward’s method iteratively merges the pair of clusters $(\mathcal{A}_a, \mathcal{A}_b)$ that minimizes the increase in within-cluster squared error,
\[
\Delta_{a,b}
= \frac{|\mathcal{A}_a|\,|\mathcal{A}_b|}{|\mathcal{A}_a| + |\mathcal{A}_b|}
\left\lVert \mu_a - \mu_b \right\rVert_2^2,
\]
until exactly $m-m'$ clusters remain, yielding a partition \\
$\{\mathcal{A}_1,\dots,\mathcal{A}_{m-m'}\}$. The pooled token embeddings are then defined as the mean of each cluster:
\[
c_j = \frac{1}{|\mathcal{A}_j|} \sum_{i \in \mathcal{A}_j} x_i, \quad j = 1,\dots,m-m',
\]

In implementation, we provide the option to keep $m'$ tokens as protected tokens and concatenate them back to the pooled token embeddings to form the final compressed sequence $\mathbf{C} = [c_1,\dots,c_m]$.

%% file: sections/50-method.tex
We now describe Attention-Guided Clustering (\ours), a compression technique designed to maximize the utility of a fixed token budget for document compression in any modality. \ours (shown in \autoref{fig:main}~(a)) combines three main components: (i) \textit{Attention-based Centroid Selection}, which utilizes learned universal query tokens to identify semantically salient information; (ii) \textit{Hard Clustering}, which uses hard assignment to group tokens to reduce redundancy while preserving distinct semantic details; and (iii) \textit{Weighted Aggregation}, which constructs the final compressed representations by averaging tokens within each cluster weighted by their saliency to mitigate the optimization challenges of hard operations.

\subsection{Attention-based Centroid Selection}
The first component of our approach is to identify 
information-rich regions within a document. To estimate token importance without specific user queries, we introduce learned ``universal queries,'' special tokens that probe the document for significant content.

Formally, we append a set of trainable tokens $\mathbf{X}_{\mathbf{\Psi}} \in \mathbb{R}^{|\mathbf{\Psi}| \times h}$, where $\mathbf{\Psi}$ denotes the set of indices for these tokens, to the document tokens $\mathbf{X}$ and pass the concatenated sequence into the bidirectional encoder $F_\text{enc}$, an $L$-layer transformer:

\begin{align*}
[\mathbf{Z}^{(L)}_\mathbf{X}, \mathbf{Z}^{(L)}_\mathbf{\Psi}] &= F_\text{enc}([\mathbf{X}, \mathbf{X}_{\mathbf{\Psi}}]; \theta)
\end{align*}

\noindent where $\theta$ represents the encoder parameters and $\mathbf{Z}^{(L)}$ denotes the hidden states at the last layer.

We then leverage the attention mechanism to quantify the importance of each document token. Let $\text{Attn}^{(L,\eta)}_{i} \in \mathbb{R}^{n}$ denote the attention weights from the universal query token $i \in \mathbf{\Psi}$ to all document tokens at the last layer $L$ and head $\eta$. To obtain a global measure of importance, we average over heads and across universal query tokens to compute the saliency scores $\boldsymbol{\alpha} \in \mathbb{R}^{n}$:

\begin{align*}
\boldsymbol{\alpha}
= \frac{1}{|\mathbf{\Psi}|\,H} \sum_{i \in \mathbf{\Psi}} \sum_{\eta=1}^{H}
\text{Attn}^{(L,\eta)}_{i}.
\end{align*}

The aim of $\boldsymbol{\alpha}$ is to capture high-level semantic relevance, allowing the model to distinguish signal from noise before clustering begins. Using these saliency scores, we then select cluster centroids. We select the top-$m$ tokens with the highest saliency scores, where $m$ is the target budget. Let $\mathcal{I} \subset \{1, \dots, n\}$ denote the indices corresponding to the $m$ largest values in $\boldsymbol{\alpha}$. We define the cluster centroids as $\mathcal{M} = \{ \boldsymbol{\mu}_k \}_{k=1}^m$, where each centroid corresponds to a selected token representation $\boldsymbol{\mu}_k = \mathbf{Z}^{(L)}_{\mathbf{X}, j}$ for some $j \in \mathcal{I}$.

\subsection{Clustering}
Next, we need to organize the rest of the tokens. We group every other token with the centroid it is most similar to, effectively gathering related context into coherent clusters. This ensures that even if a word isn't selected as a centroid, its information is preserved by being associated with a relevant cluster.

Formally, we assign every document token to its nearest centroid based on cosine similarity:

\begin{align*}
  \mathcal{G}_k = \left\{ j \in \{1, \dots, n\} \;\middle|\; k = \operatorname*{argmax}_{k' \in \{1, \dots, m\}} \cos(\mathbf{Z}^{(L)}_{\mathbf{X}, j}, \boldsymbol{\mu}_{k'}) \right\}.
\end{align*}

Similar to \hcluster, this clustering step reduces redundancy by grouping semantically similar tokens. However, unlike \hcluster, which relies on iterative agglomerative merging, our approach anchors clusters around the globally salient centroids identified by the universal queries. This ensures that the compression is guided by semantic importance rather than just local geometric proximity. Furthermore, by employing hard assignment rather than fully soft operations (e.g., \mem), we ensure that distinct semantic concepts remain separated in the latent space, alleviating the risk of over-smoothing. 

\subsection{Weighted Aggregation}
\label{subsec:weighted_aggregation}
With clusters established, we aggregate the tokens in each group into a compact representation. Naive averaging treats all inputs equally, ignoring the varying information density typical of multimodal data. Just as P-frames in video are compressed more heavily than I-frames \cite{legall1991mpegvideocompressionstandard}, or text outweighs margins in a document, we must distinguish signal from redundancy. We therefore employ \textit{Weighted Aggregation}, where the saliency score $\boldsymbol{\alpha}$ naturally serves as a learnable importance weight to prioritize critical content.

Formally, we construct the document's compressed representation $\mathbf{C} = [\mathbf{c}_1, \dots, \mathbf{c}_m] \in \mathbb{R}^{m \times h}$. We compute each cluster vector $\mathbf{c}_k$ as the weighted average of the document tokens assigned to it:

\begin{align*}
\mathbf{c}_k = \frac{\sum_{j \in \mathcal{G}_k} \boldsymbol{\alpha}_j \mathbf{Z}^{(L)}_{\mathbf{X}, j}}{\sum_{j \in \mathcal{G}_k} \boldsymbol{\alpha}_j}.
\end{align*}

This also ensures that while the structure is discrete, the contribution of each token remains continuous, allowing gradients to flow back to the feature encoder and capture fine-grained semantic variations.

%% file: sections/60-experiments.tex
\input{tables/00-all-results}

\subsection{Datasets}
We evaluate the multi-vector compression methods on several datasets spanning text, image, and video. 
\begin{itemize}
    \item \beir \cite[text,][]{thakur2021beirheterogenousbenchmarkzeroshot} is a collection of text retrieval tasks. 
    For our evaluation we chose the set of publicly available datasets with corpora of fewer than 1M documents. 
    We further exclude Quora, as it is a duplicate retrieval task in which the "documents" are duplicate questions, and therefore not in need or amenable to compression. 
    The remaining datasets span medical, financial, and argument domains.
    
    \item \vidore \textsc{v2} \cite[visual document,][]{mace2025vidorebenchmarkv2raising} is a visual document retrieval benchmark designed to evaluate systems on visually rich PDFs where information is conveyed through both text and layout (e.g., figures, tables). It consists of four datasets spanning insurance, biomedical, economics, and ESG domains, featuring long-form and cross-document queries that require multimodal understanding.
    \item \msrvtt \cite[vision-only video,][]{xu2016msrvttlargevideodescription} is a video captioning dataset converted to text-to-video retrieval, where each query is a sentence description of a video. There is only one relevant video per query and 1000 query-video pairs in test with no additional irrelevant videos.  
    \item \multivent \cite[audiovisual video,][]{kriz2025multivent20massivemultilingual} is a text-to-video retrieval dataset with queries that target visual and audio information. There are ten relevant videos per query with 2546 queries and 109,800 videos in test.
\end{itemize}

\subsection{Experimental Setup}

\paragraph{\beir}
We begin finetuning from an already finetuned single-vector encoder \cite{li2023generaltextembeddingsmultistage, zhang2024mgtegeneralizedlongcontexttext}.\footnote{\texttt{Alibaba-NLP/gte-modernbert-base}}
We train for 10,000 steps with a distillation loss on 16-way MSMARCO \cite{bajaj2018msmarcohumangenerated} hard negatives scored by a reranker.\footnote{\texttt{lightonai/ms-marco-en-bge-gemma}}
We train with a batch size of 20, learning rate of $10^{-4}$, and bfloat16 precision. In both the training and evaluation settings, we use a maximum query length of 32 (with the usual ColBERT-style query augmentation mechanism), maximum document length of 300, and no query/document marker tokens.
For retrieval, document embeddings are indexed and retrieved in a FastPlaid \cite{fastplaid2025, santhanam2022plaidefficientenginelate} index with 4-bit residuals.

\paragraph{\vidore \textsc{v2}} 
We enable bidirectional attention and initialize the pretrained weights from \textsc{Qwen2.5-VL-3B}. We train on the ColPali train set\footnote{\texttt{vidore/colpali\_train\_set}} for 2 epochs with a global batch size of 112 and gradient accumulation step of 4, learning rate of $10^{-5}$, and bfloat16 precision. We prepend "Passage: " and "Query: " for document and query respectively. Due to the combination of the large embedding dimension (2048) and quantity of embeddings ($>1000$ per document, uncompressed), we cannot fit our full uncompressed data in a FastPlaid index, and therefore resort to a brute-force search over a flat index. For the compressed methods, we are able to use a FastPlaid index, but we continue to use the flat index for compression methods to fairly compare with the baselines.

\paragraph{\msrvtt}
We again enable bidirectional attention and initialize the pretrained weights from \textsc{Qwen2.5-VL-3B}, \textsc{Qwen2.5-VL-7B}, and \textsc{Qwen3-VL-4B} for different variants. We use a fixed number of frames of 24. We train on the \msrvtt Train 9k split for 2 epochs with a global batch size of 28 and gradient accumulation step of 4, learning rate of $10^{-5}$, and bfloat16 precision. We build a flat index of each method to gather the matching positions and strengths for analysis in \autoref{subsec:util}.

\paragraph{\multivent}
We enable bidirectional attention and initialize the pretrained weights from \textsc{Qwen2.5-Omni-3B}. We train on a combination of the human written queries and synthetically generated queries \cite{skow2026rankvideoreasoningrerankingtexttovideo}. We sample frames at most 24 frames and audio at 4KHz. We train for 2 epochs with a global batch size of 8 and gradient accumulation step of 4, learning rate of $10^{-5}$, and bfloat16 precision. We cannot build any index over the uncompressed representations, and only utilize FastPlaid for the compressed representations. 

\paragraph{Evaluation}
For each dataset, we report appropriate recall at k (R@k) and normalized discounted cumulative gain at k (nDCG@k). We also report the percentage of base performance for each compression method calculated as $(\frac{\text{compression score}}{\text{base score}})$ and report the compression ratio as $1-(\frac{\text{budget}}{\text{avg toks per doc}})$.

\subsection{Results}
In \autoref{tab:all-results}, we summarize the retrieval performance of the index compression methods in retrieval settings across each modality. Our main finding is that \ours performs the best across the modalities compared to the other compression techniques, maintaining 97\% of the uncompressed model performance at nDCG@10. We also find that \hcluster performs well for non-parametric compression, often outperforming the other learned methods \seq and \mem on non-text benchmarks.  

Across the datasets, we find that the only method able to outperform the base model, which builds a full index with a one-to-one mapping between document tokens and vectors, is \ours (R@1 on \msrvtt). This highlights two main takeaways. (1)~Training multimodal retrieval methods with a compression objective can be beneficial. Multimodal (audio and visual) tokens do not always provide new information to the document representations and are often semantically redundant\footnote{Visual compression has traditionally relied on this trend \cite{barlow2001redundancyreductionrevisited}.}, meaning that information density does not scale linearly with document length, an assumption reasonably held in text. (2)~The base model underutilizes its full document representation. In \autoref{subsec:util}, we show that the base model only utilizes about $\sim 1\%$ of the representations. Practical results imply that full indices for multimodal collections offer diminishing returns relative to their cost.

\input{tables/00-beir-main}
\paragraph{\beir}
In \autoref{tab:beir-main}, we explore the methods' performance for text retrieval on a subset of \beir datasets, reporting both the absolute nDCG@10 and relative performance compared to the uncompressed ColBERT baseline.
With a budget of 32 tokens, documents in \beir are compressed by around 80\%.\footnote{NFCorpus: 87\% (avg 237 toks), FiQA: 76\% (avg 134 toks), SciFact: 86\% (avg 230 toks), SciDocs: 83\% (avg 188 toks), TREC-COVID: 81\% (avg 170 toks), Touche: 79\% (avg 153 toks), Arguana: 82\% (avg 177 toks)} We find that there is not much difference between \mem and \ours. Both methods compress the text document representations well and maintain stable performance across the different datasets. Additionally, we find that there is a larger gap between \hcluster and the learned methods, as the performance varies more significantly depending on the task.

\paragraph{\vidore}
In \autoref{tab:vidore-breakdown-updated}, we break down the performance of each method on the \vidore topic splits. Comparing the runs with the same training configurations, we see that \ours and \hcluster significantly outperform \seq and \mem. \hcluster and \ours appear to be relatively equivalent, with their averages only differing by 0.002 in nDCG@5. However, when looking at the breakdown by topic, we see that \ours is more stable across domains than \hcluster. We also compare to another learned compression method, \textsc{MetaEmbed}~\cite{xiao2025metaembedscalingmultimodalretrieval}. We find that \ours and \hcluster have comparable or better performance to \textsc{MetaEmbed}. This highlights the strengths of both \ours and \hcluster, even when training at smaller scales.\footnote{We note that the comparison to \textsc{MetaEmbed} is not 1-to-1 because we are only capable of training at $\frac{1}{20}$th of the scale.}

\input{tables/00-vidore-main}

\paragraph{\msrvtt}
In \autoref{tab:msrvtt-main}, we report more detailed comparisons of our baseline and method on \msrvtt. We find that every compression method sets a new state-of-the-art on MSR-VTT over previous multi-vector approaches (\textsc{ColQwen-Omni}, \textsc{Video-ColBERT}) and dense approaches (\textsc{OmniEmbed}), even when compressing the index to only 5 vectors per document. We even see at budgets of 32 and 128, \ours performs better at R@1 than the base model, again showing that training for compression in multimodal retrieval may not only lead to efficient indices, but also best performance. 

\input{tables/00-msrvtt-main}

\paragraph{\multivent} 
In \autoref{tab:all-results}, we report the retrieval results for \multivent on the compression methods. We are not able to build the index for the full model as comparable vision indices use a hidden dimension of 2048 \cite{reddy2025videocolbertcontextualizedlateinteraction, xu2025llamanemoretrievercolembedtopperforming, moreira2026nemotroncolembedv2topperforming, xiao2025metaembedscalingmultimodalretrieval}, demonstrating that for large multimodal indices compression is necessary. Unlike \vidore and \msrvtt, evaluating on \multivent requires leveraging the audio information to retrieve the relevant videos. We found inefficient audio sampling to be a major limitation of the \textsc{Qwen-Omni} model, suggesting interesting future work on how to pass audiovisual signal efficiently to MLLMs. For example, when training on the \multivent data, we found that in order to fit a batch size of 8, the audio signal needed to be reduced from 16KHz (\textsc{Qwen-Omni}'s training rate) to 4KHz to fit on the devices.\footnote{Sampling audio below 4KHz degrades speech intelligibility \cite{baer2002effectslowpassfiltering}.} 

\input{tables/11-budget}
\input{tables/10-stability}

\subsection{Compression Ranges and Stability}
\label{subsec:stability}
\paragraph{Compression Ranges}
In \autoref{tab:msrvtt-main}, we explore learning different ranges of compression on MSR-VTT for 5 tokens (99.62\%), 32 tokens (97.57\%) and 128 tokens (90.29\%). For \seq, we see that performance seems to be relatively flat across the compression ratios, with similar R@1 and nDCG@10 results. This is an interesting finding, largely suggesting that \seq may underutilize the budget and lead to a suboptimal index (see further evidence in \autoref{subsec:util}). For all other methods, we see an increase in performance from the most extreme compression to lighter ratios. Additionally, we find \hcluster's performance impressive as a non-parametric technique at a budget of 5. However, \ours continues to have strongest performance at each ratio, demonstrating its robustness to a variety of compression ratios. 

\paragraph{Budget Sweeps}
In \autoref{tab:budget}, we analyze the impact of varying token budgets and the number of appending tokens of \ours on retrieval performance on \msrvtt. We observe that performance scales positively with both the size of the token budget and the quantity of appending tokens. Notably, even under the most extreme compression setting (a budget of 5), \ours maintains robust performance, outperforming the single dense vector encoder, OmniEmbed-7B, despite using a smaller 3B backbone. When looking at the number of appended query tokens and the budget, we find that it is generally optimal to align the number of appended query tokens and the budget size. Additionally, we see that 32 appended query tokens and a budget of 5 outperforms 5 appended query tokens at the same budget, but we don't see the same pattern for 128 appended query tokens and 32 budget. This suggests that it is important to avoid a low number of query tokens, but that performance doesn't necessarily scale to the number of appended query tokens at any budget. 

\input{tables/12-generalize}

\paragraph{Compression Transferability}
In \autoref{tab:stability}, we again explore different ranges of compression on MSR-VTT, but only train the model for a single compression ratio (i.e., training \ours for 32). We find that \ours provides the best ability to generalize to an unseen compression ratio after training. We attribute this finding to a weakness in the heuristic of \hcluster, redundant visual tokens should not be equally merged. By leveraging attention to select centroids and weight the merging, \ours is better able to preserve salient semantic concepts and reduce the redundancy along the temporal dimension. We do not compare to \mem and \seq in this experiment as they do not generalize to new compression ratios. 

Looking closer at the \ours results, we observe very close performance between methods trained for budgets of 5 and 128 and the method trained for 32 and tested on those budgets. This demonstrates that our model can transfer abilities between budgets at performance near training \ours for that compression ratio.    

\paragraph{Model Size and Backbone}
In \autoref{tab:model_generalizability}, we also provide an experiment on the stability of our method with a different model size and with a different underlying model. When scaling \ours to a 7B parameter model, initializing from \textsc{Qwen2.5-VL-7B}, we find that performance is significantly improved. Additionally, we initialize our model from \textsc{Qwen3-VL-4B} and find that performance again improves over both Qwen2.5 versions. These results demonstrate that our method benefits from models with stronger representations and stronger encoding abilities, and should scale to any backbone or model size.

\input{figures/utilization_fig}
\input{figures/utilization_fig2}
\subsection{Index Utilization}
\label{subsec:util}
In \autoref{fig:utilization}, we visualize the difference in index utilization for full uncompressed indices and the four compression techniques and the similarity between document vectors in each index on \msrvtt. To calculate the matching strength, we sum the maximum similarity scores across all relevant query-document pairs in \msrvtt and normalize by the total number of match records at that query position. For the heatmaps, we calculate the cosine similarity between token vectors within each document, averaged across documents in the index of \msrvtt.

\paragraph{Token Utilization Analysis} Our analysis of index statistics reveals that of the 1.3 million unique document tokens, only $\sim 1\%$ are active during a single evaluation pass, with the base model primarily utilizing the first 2\%. 
This again stresses that building full indices for multimodal collections is unnecessary and that these indices can greatly benefit from compression. For \seq, we see a unique trend amongst the compression methods, only selecting a few tokens from the document representation in late interaction. This underutilization of the budget corroborates the findings of \autoref{tab:msrvtt-main}, where \seq's performance seemed to plateau across the compression ratios. For \mem and \ours, we see that both methods attempt to utilize their full document representations. Because \mem's representations are appended to the document context, we see a significant bias towards the first few tokens in the representation. This result largely follows the trend in dense encoding with causal language models to append a token at the end of the sequence to use as the document representation \cite{ma2024finetuningllamamultistagetext, liu2024llama2vecunsupervisedadaptationlarge}. Unlike \mem, \ours and \hcluster use representations from the document leading to a better utilization of the compressed representations in late interaction.

\paragraph{Token Similarity Analysis} In the heatmaps of \autoref{fig:utilization}, we further investigate the internal structure of these indices by visualizing the cosine similarity between document tokens. For the full model, we observe that the first few tokens, which we previously noted dominate late interaction, exhibit a consistently high similarity ($\sim 0.7$) to nearly all other tokens in the document. Additionally, these initial tokens are highly similar to one another. This similarity explains the significant imbalance in matching strength observed in the corresponding bar chart. \seq presents a distinct pattern where tokens that are never used in interaction display negative similarity. We interpret this as a modeling failure; tokens derived from the same document context should theoretically maintain a baseline degree of positive similarity, which \seq fails to capture. Conversely, \mem demonstrates an over-smoothing problem, where the heatmap is dominated by high similarity scores. This lack of diversity restricts the expressive power of the index, as the tokens fail to capture distinct semantic nuances. \hcluster, by design, merges similar tokens and consequently produces the most diverse set of representations, as evidenced by the lower off-diagonal similarities. However, this suggests that similarity-based heuristics alone are not sufficient for optimal performance, as \hcluster does not perform as well as learned methods in many settings despite its high diversity. Finally, \ours shows a trend similar to \hcluster but maintains decent inter-token similarities. This balance highlights the efficacy of our approach: it avoids the representation collapse seen in \mem while preserving necessary semantic overlaps that \hcluster lacks, resulting in a robust compressed index.

\input{tables/correlation}
\paragraph{Predicting Performance with Utilization}

Following from the above observations, we explore if it is possible to predict compression performance by only looking at how evenly distributed the strength of maximum similarity matches are in a document representation. We calculate the Coefficient of Variation (CV) and the Gini coefficient. The CV assesses relative variability standardized by the mean $CV = \frac{\sigma}{\mu} \times 100$, while the Gini coefficient quantifies distributional concentration $G = \frac{2 \sum i x_i}{n \sum x_i} - \frac{n+1}{n}$. For both metrics, lower values indicate a more evenly distributed activations.

In \autoref{tab:correlation} and \autoref{fig:correlation_plot}, we show the Pearson's correlation \cite{PearsonVIINO} between the retrieval metrics (R@1, nDCG@10, and MRR) and the inverse evenness metrics (Coefficient of Variation (CV) and Gini Coefficient).\footnote{Because of the difficulty in obtaining different indices, we compute these correlations with 5 samples. Larger scale exploration is needed to further validate this finding.} We find a rough correlation between the evenness of the distribution of maximum similarity matches and retrieval performance. These results suggest that training late interaction methods to maximize the utility of each token in its document representations will lead to strong performance, which we leave for future work to explore. Additionally, this finding suggests that during development it is satisfactory to estimate the downstream performance of a compression method with the distribution of maximum similarity matches on a small set of queries. This is especially beneficial for multimodal tasks, where building indices is storage and time expensive.  

\subsection{Method Ablation}

In \autoref{tab:ablation}, we perform an ablation analysis on the modeling choices in \ours using \msrvtt. We examine the contribution of three components: Attention-based Centroid Selection, Clustering, and Weighted Aggregation. First, removing the attention weights (w/o Attn Weight) from the aggregation step leads to a decline in performance. As discussed in \autoref{subsec:weighted_aggregation}, weighting the contribution of tokens by their saliency scores is beneficial for balancing the hard assignment operation with optimization stability. Without these weights, the contribution of individual tokens becomes less continuous, rendering the optimization landscape rougher and less effective. Second, we analyze the impact of attention-based selection (w/o Attn Select) by replacing the learned universal query tokens with a random selection strategy. This prevents the model from distinguishing signal from noise. While this randomness helps maintain some diversity, it forces the model to compress mixed information into each vector. This saturates the capacity of each vector and weakens its ability to capture complex, discriminative semantics. Finally, we evaluate the model without clustering (w/o Cluster), relying solely on attention selection. Without the clustering operation to reduce redundancy and aggregate context, the resulting representations lack diversity and expressiveness. Consequently, token matches in one evaluation pass become highly concentrated on a narrow set of tokens, leading to worse performance. As shown in \autoref{fig:correlation_plot}, the progressive removal of Weighted Aggregation and Clustering leads to a simultaneous decline in retrieval performance and evenness of MaxSim matches. This confirms that \ours effectively incentivizes balanced utilization while improving retrieval.

\input{tables/ablation}

%% file: tables/00-all-results.tex
\begin{table*}[ht]
    \centering
    \begin{tabular}{l|cccccccc}
    \toprule
        \multirow{2}{*}{\textbf{Method}} & \multicolumn{2}{c}{\textbf{\beir}} & \multicolumn{2}{c}{\textbf{\vidore}} & \multicolumn{2}{c}{\textbf{\msrvtt}} & \multicolumn{2}{c}{\textbf{\multivent}} \\ 
        & R@10 & nDCG@10 & R@1 & nDCG@5 & R@1 & nDCG@10 & R@10 & nDCG@10 \\
    \midrule
        Baseline 
            & 37.1 & 46.2 
            & 27.7 & 60.0
            & 55.7 & 71.9
            & --$^*$ & --$^*$ \\
    \midrule
        \multirow{2}{*}{\seq}
            & 35.8 & \underline{43.9}
            & 23.5 & 51.7
            & 53.3 & 69.9
            & 41.1 & 38.5 \\
            
            & 96.5\% & 95.0\% 
            & 84.8\% & 86.2\% 
            & 95.7\% & 96.9\% 
            & N/A & N/A \\
            
        \multirow{2}{*}{\mem}
            & \underline{36.3} & \textbf{45.0}
            & 25.0 & 54.4
            & \underline{54.2} & 69.9
            & 48.7 & 44.8\\

            & 97.8\% & 97.4\% 
            & 90.3\% & 90.7\% 
            & 97.3\% & 96.9\% 
            & N/A & N/A \\
            
        \multirow{2}{*}{\hcluster}
            & 35.5 & 41.2
            & \underline{26.0} & \underline{56.4} 
            & 54.1 & \underline{70.4}
            & \underline{49.2} & \textbf{46.5} \\
            
            & 95.7\% & 89.2\% 
            & 93.9\% & 94.0\%
            & 97.1\% & 97.6\% 
            & N/A & N/A \\
    \midrule 
        \multirow{2}{*}{\ours (Ours)}
            & \textbf{37.0} & \textbf{45.0} 
            & \textbf{26.3} & \textbf{56.7} 
            & \textbf{56.9} & \textbf{71.5}
            & \textbf{49.6} & \underline{46.3} \\

            & 99.7\% & 97.4\% 
            & 94.9\% & 94.5\% 
            & 102.1\% & 99.2\%
            & N/A & N/A \\
    \bottomrule
    \end{tabular}
    \caption{Results of index compression on each retrieval benchmark. Compression budgets: \beir 32, \vidore 64, \msrvtt 32, \multivent 64. $*$ means baseline was unable to build due to compute. Second row of each method shows percent of baseline.}
    \label{tab:all-results}
    \vspace{-1em}
\end{table*}

%% file: tables/00-beir-main.tex
\begin{table}[tbp]
\setlength{\tabcolsep}{3.5pt}
    \centering
    \begin{tabular}{c|c|ccccccc}
    \toprule
        \textbf{Meth.} & \textbf{Avg} & \textbf{NF} & \textbf{FQA} & \textbf{SciF} & \textbf{SciD} & \textbf{TC} & \textbf{Tou} & \textbf{Arg}\\

    \midrule
        Base & 46.2 & 37.1 & 43.3 & 74.4 & 18.6 & 78.5 & 27.5 & 44.1 \\
    \midrule
        \multirow{2}{*}{\textsc{Seq}}
            & 43.9 & 31.0 & 36.8 & 70.3 & 18.7 & 75.8 & 24.8 & 50.0 \\
            & \underline{95.0} & 83.6 & 85.0 & 94.5 & \textbf{100.5} & 96.6 & \underline{90.2} & \textbf{113.4} \\ 
    \midrule
        \multirow{2}{*}{\textsc{Mem}}
            & 45.0 & 35.2 & 39.8 & 71.7 & 17.6 & 77.6 & 27.7 & 45.1\\
            & \textbf{97.4} & \underline{94.9} & \textbf{91.9} & 96.4 & 94.6 & \underline{98.9} & \textbf{100.7} & 102.3\\
    \midrule
        \multirow{2}{*}{\textsc{H-P}}
            & 41.2 & 33.7 & 36.2 & 72.5 & 18.4 & 62.0 & 17.6 & 48.3\\
            & 89.1 & 90.8 & 83.6 & \textbf{97.4} & \underline{98.9} & 79.0 & 64.0 & \underline{109.5}\\
    \midrule 
        \multirow{2}{*}{\ours}
            & 45.0 & 36.0 & 38.9 & 72.4 & 18.1 & 78.7 & 23.3 & 47.6\\
            & \textbf{97.4} & \textbf{97.0} & \underline{89.8} & \underline{97.3} & 97.3 & \textbf{100.3} & 84.7 & 107.9\\
    \bottomrule
    \end{tabular}
    \caption{nDCG@10 (top) and percent performance relative to uncompressed baseline (bottom) on \beir datasets.  NF: NFCorpus, FQA: FiQA, SciF: SciFact, SciD: SciDocs, TC: TREC-COVID, Tou: Touche, Arg: arguana.}
    \label{tab:beir-main}
    \vspace{-2em}
\end{table}

%% file: tables/00-vidore-main.tex
\begin{table}[tbp]
  \centering
  \begin{tabular}{lc|c|cccc}
  \toprule
      \textbf{Method} & \textbf{Tok} & \textbf{Avg} & \textbf{Bio} & \textbf{Econ} & \textbf{ESG-R} & \textbf{ESG-H} \\
  \midrule
      Base & 1297 & 60.0 & 61.4 & 53.9 & 57.0 & 67.6 \\
      ColPali & - & 53.3 & 56.5 & 49.9 & 55.7 & 51.1 \\
      CQO & - & 56.5 & 56.5 & 53.2 & 54.2 & 62.2 \\
      \textsc{MEmbed} & 64 & 58.8 & 58.7 & 55.5 & 57.4 & 63.7 \\
  \midrule
      \seq & 64 
          & 51.7 
          & 54.7 
          & \underline{53.5} 
          & 45.2 
          & 53.5 \\
      \mem & 64
          & 54.3 
          & 56.8 
          & 53.0 
          & 46.4 
          & \textbf{61.4} \\
      
      \hcluster & 64 
          & \underline{56.4} 
          & \textbf{59.6} 
          & 52.1 
          & \underline{53.4} 
          & \underline{60.6} \\
      \ours & 64
          & \textbf{56.7} 
          & \underline{59.0} 
          & \textbf{54.5}
          & \textbf{55.8} 
          & 57.3 \\
  \bottomrule
  \end{tabular}
  \caption{nDCG@5 breakdown by domain on \vidore \textsc{v2} for the multilingual subsets. Bio: Biomedical, Econ: Economics, ESG-R: ESG Reports, ESG-H: ESG Human. ESGs are English. CQO: ColQwenOmni, MEmbed: MetaEmbed.}
  \label{tab:vidore-breakdown-updated}
  \vspace{-1em}
\end{table}

%% file: tables/00-msrvtt-main.tex
\begin{table}[tbp]
    \centering
    \begin{tabular}{cc|ccccc}
    \toprule
        \textbf{Tok} & \textbf{Method} & \textbf{R@1} & \textbf{R@10} & \textbf{nDCG@10} \\
    \midrule
        1 & OmniEmbed-7B \cite{ma2025tevatron20unifieddocument} & 51.5 & 83.2 & 67.1  \\
        26 & Video-ColBERT \cite{reddy2025videocolbertcontextualizedlateinteraction} &  51.5 & 85.5 & 67.7 \\
        1702 & ColQwen-Omni 3B & 40.8 & 73.8 & 56.3 \\
        1318 & Baseline 3B (Ours) & \textbf{55.7} & \textbf{88.3} & \textbf{71.9} \\
    \midrule
        \multirow{4}{*}{5}
            & \seq & \underline{53.4} & \underline{86.7} & \textbf{69.5} \\
            & \mem & 52.7 & \textbf{87.3} & \underline{69.3}\\ 
            & \hcluster & 52.6 & 86.1 & 68.9 \\
            & \ours & \textbf{53.9} & 85.8 & 69.2 \\
    \midrule
        \multirow{4}{*}{32}
            & \seq & 53.3 & 86.9 & 69.9 \\
            & \mem & \underline{54.2} & 86.4 & 69.9 \\ 
            & \hcluster & 54.1 & \textbf{87.3} & 70.4 \\
            & \ours & \textbf{56.9} & \underline{87.0} & \textbf{71.5} \\
    \midrule
        \multirow{4}{*}{128}
            & \seq & 52.6 & \textbf{87.6} & 69.7 \\
            & \mem & \underline{54.9} & 86.7 & 70.5 \\ 
            & \hcluster & 54.4 & \underline{87.2} & \underline{70.9} \\
            & \ours & \textbf{56.4} & \textbf{87.6} & \textbf{71.6} \\
    \bottomrule
    \end{tabular}
    \caption{Retrieval performance on \msrvtt. We compare compressed methods against uncompressed baselines at budgets of 5, 32, and 128 tokens, alongside SOTA models.}
    \label{tab:msrvtt-main}
    \vspace{-1em}
\end{table}

%% file: tables/11-budget.tex
\begin{table}[ht]
  \centering
  \begin{tabular}{cc|ccc}
  \toprule
  \textbf{Tok} & \textbf{Appn Tok} & \textbf{R@1} & \textbf{R@10} & \textbf{nDCG@10} \\ 
  \midrule
  \multirow{2}{*}{5} & 5 & 53.9 & 85.8 & 69.2 \\
    & 32 & 53.1 & 86.7 & 69.6 \\
  \midrule
  \multirow{3}{*}{32} & 5 & 54.2 & 87.2 & 70.6 \\
    & 32 & 56.9 & 87.0 & 71.5 \\
    & 128 & 55.4 & 87.6 & 71.2 \\
  \midrule
  \multirow{2}{*}{128} & 32 & 55.2 & 86.8 & 70.8 \\
    & 128 & 56.4 & 87.6 & 71.6 \\
  \bottomrule
  \end{tabular}
  \caption{Retrieval performance metrics of AGC across varying budgets and appending tokens on MSR-VTT. Each combination is used with the same configurations of training. Appn Tok: number of appending tokens.}
  \label{tab:budget}
  \vspace{-2em}
\end{table}

%% file: tables/10-stability.tex
\begin{table}[!t]
    \begin{tabular}{l|cc|ccc}
      \toprule
      \textbf{Method} & \textbf{Train} & \textbf{Test} & \textbf{R@1} & \textbf{R@10} & \textbf{nDCG@10} \\
      \midrule
      Baseline 
          & 1318 & 1318 
          & 55.7 & 88.3 & 71.9 \\
      \midrule
      \multirow{3}{*}{\ours}
          & 32 & 5 
          & \textbf{53.6} & \textbf{87.4} & \textbf{70.1} \\
          & 32 & 32 
          & \textbf{56.9} & 87.0 & \textbf{71.5} \\
          & 32 & 128 
          & \textbf{56.4} & \textbf{87.5} & \textbf{71.7} \\
      \midrule
      \multirow{3}{*}{\hcluster}
          & 1318 & 5 
          & 52.6 & 86.1 & 68.9 \\
          & 1318 & 32 
          & 54.1 & \textbf{87.3} & 70.4 \\
          & 1318 & 128 
          & 54.4 & 87.2 & 70.9 \\
      \bottomrule
    \end{tabular}
    
    \caption{Generalizability of \ours and \hcluster compression methods on \msrvtt.}
    \label{tab:stability}
    \vspace{-1em}
  \end{table}

%% file: tables/12-generalize.tex
\begin{table}[!t]
    \centering
    \begin{tabular}{l|ccc}
      \toprule
      \textbf{Model Variant} & \textbf{R@1} & \textbf{R@10} & \textbf{nDCG@10} \\
      \midrule
      Qwen2.5-VL-3B & 56.9 & 87.0 & 71.5 \\
      Qwen2.5-VL-7B & 58.0 & 89.0 & 73.0 \\
      Qwen3-VL-4B & 58.5 & 88.4 & 73.0 \\
      \bottomrule
    \end{tabular}
    \caption{Generalizability of \ours across different model sizes and variants. Results indicate consistent performance scaling with larger and newer model backbones.}
    \label{tab:model_generalizability}
    \vspace{-1em}
\end{table}

%% file: figures/utilization_fig.tex
\begin{figure*}
    \centering
    \includegraphics[width=\linewidth]{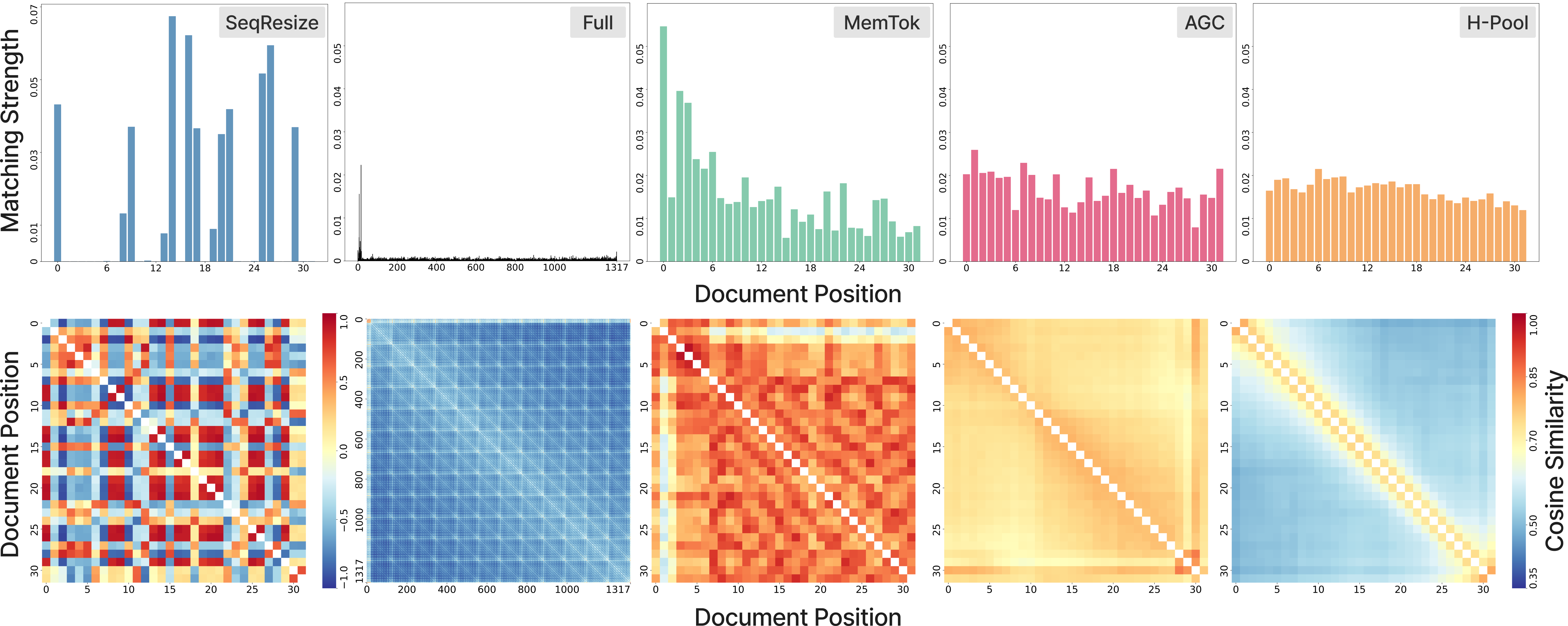}
    \vspace{-2em}
    \caption{
    Index utilization and inter-position similarity analysis on MSR-VTT. Top row: Per-position matching strength for each method, computed by summing the maximum similarity matches between query tokens and document tokens across all relevant query-document pairs, averaged over query positions. Bottom row: Pairwise cosine similarity between document vectors within each document, averaged across all documents in the index.
    }
    \label{fig:utilization}
    \vspace{-1em}
\end{figure*}

%% file: figures/utilization_fig2.tex
\begin{figure}
    \centering
    \includegraphics[width=\linewidth]{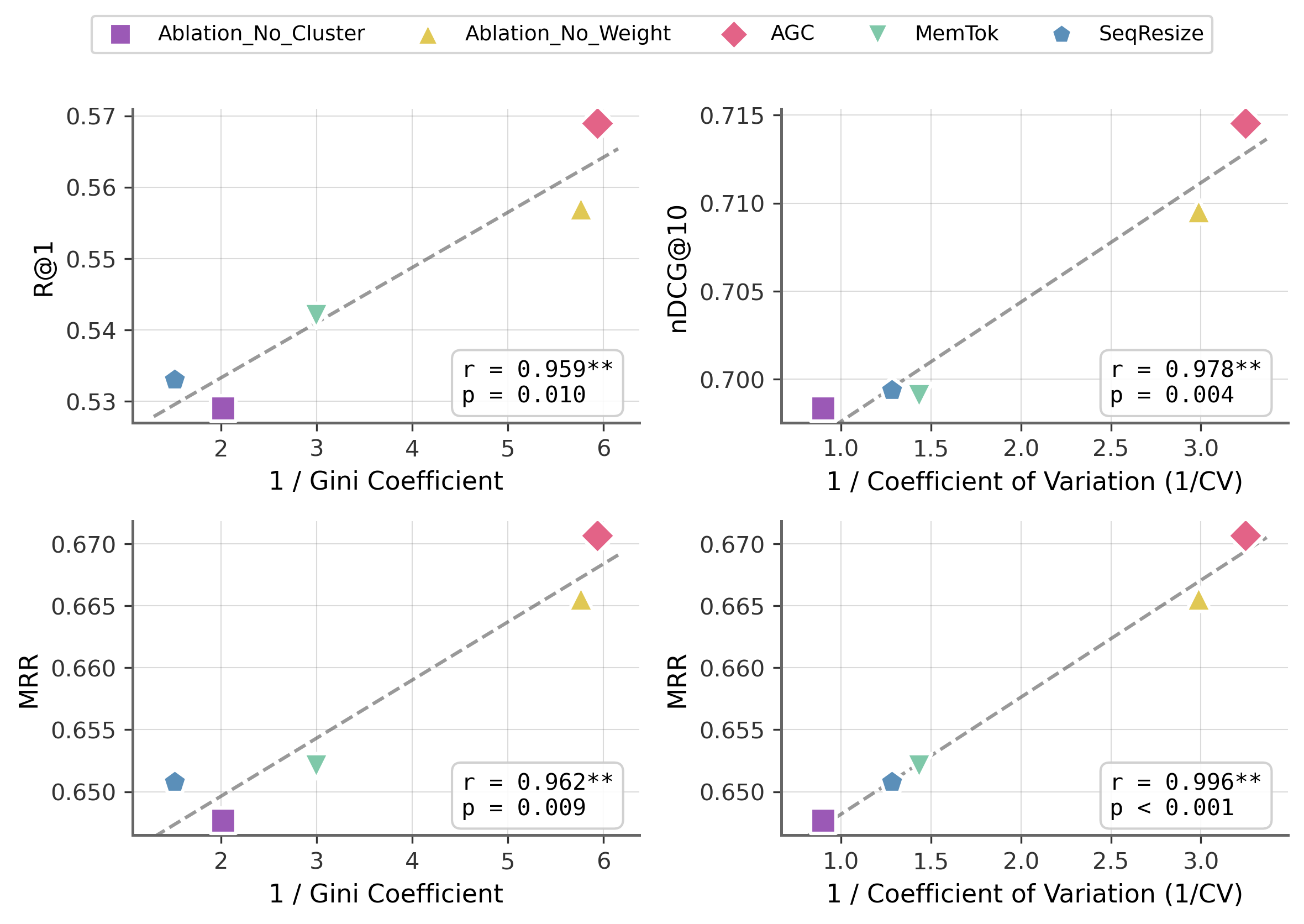}
    \caption{Correlation between retrieval performance metrics and distribution evenness measures on MSR-VTT dataset. Dashed lines indicate linear regression fits. All correlations are statistically significant ($p \le 0.01$), with Pearson's r ranging from 0.959 to 0.996.}
    \label{fig:correlation_plot}
    \vspace{-1em}
\end{figure}

%% file: tables/correlation.tex
\begin{table}[t]
\centering
\small
\begin{tabular}{l|cccc}
\toprule
    & \multicolumn{2}{c}{\textbf{CV}} & \multicolumn{2}{c}{\textbf{Gini}} \\
    \textbf{Metric} & \textbf{Pearson $r$} & \textbf{p-value} & \textbf{Pearson $r$} & \textbf{p-value} \\
\midrule
    R@1 & 0.974$^{**}$ & 0.005 & 0.959$^{**}$ & 0.010 \\
    nDCG@10 & 0.978$^{**}$ & 0.004 & 0.943$^{*}$ & 0.016 \\
    MRR & 0.996$^{**}$ & $<$0.001 & 0.962$^{**}$ & 0.009 \\
\bottomrule
\end{tabular}
\caption{Pearson correlation analysis between retrieval metrics and inverse evenness metrics (1/evenness) on \msrvtt, testing the hyperbolic relationship retrieval $\sim$ 1/evenness. All variants use a fixed budget of 32. Evenness metrics measured on matching strength of (document position, query position) pairs. CV: Coefficient of Variation, Gini: Gini Coefficient. Significance levels: $^{**} p < 0.01$, $^{*} p < 0.05$.}
\label{tab:correlation}
\vspace{-2em}
\end{table}

%% file: tables/ablation.tex
\begin{table}[t]
  \centering
  \begin{tabular}{l c c c}
  \toprule
  \textbf{Method} & \textbf{R@1} & \textbf{R@10} & \textbf{nDCG@10} \\
  \midrule
  AGC & 56.9 & 87.0 & 71.5 \\
  \midrule
  w/o Attn Weight & 55.7 & 86.5 & 71.0 \\
  w/o Attn Select & 54.1 & 86.8 & 70.0 \\
  w/o Cluster & 52.9 & 87.3 & 69.8 \\
  \bottomrule
  \end{tabular}
  \caption{Ablation Study Results on \msrvtt. We observe a drop in performance as components are removed (individually, not in sequence) from the full AGC model.}
  \label{tab:ablation}
  \vspace{-2em}
\end{table}

%% file: sections/70-conclusion.tex
In this work, we explore multi-vector index compression in any modality. We adapt three strong text-based compression methods to multimodal documents: \seq (projection-based), \mem (token-based), and \hcluster (heuristic-based), and introduce \ours, a novel approach to multi-vector compression. \ours uses three main aspects---attention-based centroid selection, clustering, and weighted aggregation---to maximize the utility of a fixed document token budget. We find that \ours is a strong and robust method to compress documents in any modality under a variety of compression ratios, domains, and model specifications. \ours consistently outperforms the other compression methods across the modalities and even sets a new state-of-the-art on \msrvtt. Finally, we visualize how each method utilizes its index, demonstrating that \ours and \hcluster properly utilize their budgets, and show that downstream retrieval performance roughly correlates to how well utilized a document representation is in late interaction. 

\paragraph{Future Work}
While our work corroborates the understanding that multi-vector embeddings can be compressed with a high compression ratio while retaining most retrieval performance, the budget is applied statically or potentially linearly in the case of \hcluster. A natural extension would be to develop compression mechanisms that allocate the budget in proportion to a document's inherent informational content, perhaps by using lightweight features like our document token utilization to calibrate the level of compression.